**Prospector: a mobile app for high-throughput NIRS phenotyping**

Trevor W. Rife*[1,+], Chaney Courtney[1,+], Jenna Hershberger[2], Brandon Shaver[3], Michael A. Gore[2], Mitchell Neilsen[3], Jesse A. Poland[1]

[1] Dept. of Plant Pathology, Kansas State University, Manhattan, KS

[2] Plant Breeding and Genetics Section, School of Integrative Plant Science, Cornell University, Ithaca, NY 14853

[3] Dept. of Computer Science, Kansas State University, Manhattan, KS

[+] These authors contributed equally to this work.

*Corresponding author

**ABSTRACT**

Quality traits are some of the most important and time-consuming phenotypes to evaluate in plant breeding programs. These traits are often evaluated late in the breeding pipeline due to their cost, resulting in the potential advancement of many lines that are not suitable for release. Near-infrared spectroscopy (NIRS) is a non-destructive tool that can rapidly increase the speed at which quality traits are evaluated. However, most spectrometers are non-portable or prohibitively expensive. Recent advancements have led to the development of consumer-targeted, inexpensive spectrometers with demonstrated potential for breeding applications. Unfortunately, the mobile applications for these spectrometers are not designed to rapidly collect organized samples at the scale necessary for breeding programs. To that end, we developed Prospector, a mobile application that connects with LinkSquare portable NIR spectrometers and allows breeders to efficiently capture NIR data. In this report, we outline the core functionality of the app and how it can easily be integrated into breeding workflows as well as the opportunities for further development. Prospector and other high throughput phenotyping tools and technologies are required for plant breeders to develop the next generation of improved varieties necessary to feed a growing global population.

**Core Ideas** [3-5 impact statements, 85 char. max for each]
- Prospector is an app for the streamlined collection of near-infrared spectral data
- Prospector connects to both models of LinkSquare spectrometers
- Data can be exported as organized CSV text files for subsequent analyses

**Abbreviations**

NIRS - Near-infrared spectroscopy

$R^2_{prediction}$ - Squared Pearson's correlation of predicted and observed values

# INTRODUCTION

With a growing population and changing global climate, plant breeders have been tasked with delivering improved crop varieties to farmers faster than ever (Ray et al., 2013). Despite considerable advances in genomics over the past several decades (Yuan et al., 2017), breeding progress remains limited by budget, labor, and phenotyping constraints (Furbank and Tester, 2011; Araus and Cairns, 2014). These challenges are especially evident when breeding targets include quality traits that require expensive and time-consuming laboratory assays. Quality traits are often not evaluated until late in the varietal development process when there are fewer samples to evaluate (Kumar et al., 2012; Battenfield et al., 2018). To facilitate efficient evaluation of breeding populations for quality traits, breeders require low-cost and high-throughput methods to screen lines in early selection stages.

Indirect measurement of quality traits via near-infrared spectroscopy (NIRS) has been used in many crops to provide a fast, reliable, and reagent-free alternative to traditional methods (Pasquini, 2018). Unfortunately, industry standard spectrometers are prohibitively expensive for many breeding programs, require extensive sample preparation before use, and are generally immobile. Recent developments in sensor technology have led to the development of several low-cost, handheld NIR spectrometers that can be used in the field with minimal sample processing, including the LinkSquare 1 and LinkSquare NIR (Stratio, Inc., San Jose, CA). These spectrometers have demonstrated predictive potential in diverse applications (You et al., 2017; Moon et al., 2020). Therefore, given the appropriate data management framework, these portable spectrometers have the potential to transform how NIRS is applied to breeding programs by allowing breeders to easily collect new traits or time series data.

A barrier to the adoption of affordable spectrometers in breeding programs is the lack of optimized software that can be used to efficiently collect data at the scale necessary for varietal development. Data are often returned to the user in a unique format requiring additional resources for conversion or processing. Manufacturer applications often have limitations related to sample name length or format and generally do not allow users to input these fields via barcodes- a staple in any efficient, modern plant breeding program.

To that end, we have developed Prospector, an open-source Android application that is capable of capturing, storing, visualizing, and exporting data from LinkSquare handheld NIR spectrometers. Prospector provides fast, reliable capture of scans for phenotype prediction with a standardized user interface and data export format developed by Hershberger et al. (2021) that has been adopted by BreedBase, a common breeding database(https://breedbase.org/). Our optimized mobile application will allow breeders to rapidly apply NIRS for phenotyping, thereby streamlining the selection process and improving plant breeding productivity.

**IMPLEMENTATION**

*Starting out*

The first screen a user sees when opening the app is an instructional guide on how to connect the wireless LinkSquare spectrometer to their Android device. If the user has just installed the app, they will also be prompted to load a sample dataset that demonstrates additional features without the need to connect the spectrometer and collect scans themselves. The primary interface is divided into two relevant sections, both accessible from the bottom toolbar: Data and Settings. Throughout the app, the spectrometer connection indicator on the right of the top toolbar will allow the user to navigate back to the connection instructions if a spectrometer is not currently paired and connected (Figure 1).

*Scanning workflow*

We designed Prospector to work around three entities common throughout plant breeding programs: experiments, samples, and scans. Experiments contain groups of samples; samples contain groups of scans. The default view of the Data section is the list of experiments the user has created. Each row in this list shows the name of an experiment, the date that an experiment was created, and the number of samples contained within that experiment (Figure 1, left). When creating a new experiment, the user inputs the experiment name and selects which LinkSquare spectrometer model they will be using for that experiment. Since the wavelength ranges between LinkSquare models differ, each experiment can only be associated with only one of the two models.

Once the user selects an experiment, they are shown a list of samples that have been created within that experiment. Each row in this list is numbered and shows the sample name, the sample creation date, and the number of scans that have been collected for that sample (Figure 1, center). Creating a new sample allows the user to assign a sample name by typing or via barcode. The top toolbar shows the experiment name, sample search, export, and the spectrometer connection status. Sample search allows the user to scan a barcode to move to a sample that already exists or quickly create a new sample. Individual samples can be deleted by swiping to the right.

Within each sample, the user is shown the individual scan data (Figure 1, right). LinkSquare devices utilize two separate light sources when scanning and each light source is graphed independently. Every captured scan is listed below the graph and visibility can be toggled to show or hide specific scans. Scans can be individually deleted by swiping right or bulk deleted with the delete option on the scan toolbar. If no spectrometer has been paired when creating a new scan, the initial connection instructions will be shown.

*Settings*

Additional Prospector functionality has been included in the Settings section of the app (Figure 2). Metadata are important in plant breeding and allows for proper data ownership and provenance. Therefore, in addition to capturing a timestamp with the creation of an experiment, sample, or scan, Prospector includes the ability to track the person collecting each individual scan. Users can also retrieve access to connected spectrometer information and change the method used for pairing the spectrometer to Prospector. Further, the amount and type of data being collected can be adjusted. Each light source used by LinkSquare devices can collect multiple "frames" within each scan. Increasing the number of frames used results in scans that take longer but with higher scan stability. Users can adjust the number of frames from zero to eight.

*Export*

Scan data are stored in an internal SQLite database. Prospector exports data in a CSV format that is supported for upload in BreedBase (Hershberger et al., 2021). The exported file contains the experiment name, the scan ID, the scan date, the device type, the device ID, the operator, the light source used for the scan, specific scan notes, and an array of columns containing the spectral data.

*Future directions*

While we believe Prospector addresses the major limitations to incorporating NIRS into breeding programs, we appreciate that breeders and breeding programs have diverse feature requirements. Integration and development of future features will depend on the feedback received from the breeding community.

One area we are currently working toward is the development and inclusion of complex data structures like NIRS scans into the Breeding API (BrAPI) (Selby et al., 2019). Incorporating this data type will ultimately allow Prospector to be rapidly deployed and tightly integrated into BrAPI-enabled databases, thus allowing breeders to quickly capture and analyze data using comprehensive spectral analysis packages (Hershberger et al., 2021).

In the longer term, Prospector has the potential to be the user-facing frontend for NIRS-based crop improvement. As additional manufacturers release new spectrometer models, we welcome opportunities to support these tools in Prospector so that they will have a positive impact in the global plant breeding community.

**CONCLUSION**

Rapid evaluation of quality traits is a key step in many breeding programs. NIRS is well-positioned to reduce many of the manual, complex, or destructive methods generally employed for phenotypic collection. To complement the availability of inexpensive, portable spectrometers, we have developed Prospector, a breeding-centric mobile application for spectral data collection and management. This app will simplify the addition of NIRS into diverse breeding workflows and help increase the rate of genetic gain across numerous crops. With Prospector, we endeavor to further develop the concept of a digital breeding ecosystem that allows rapid data acquisition, analysis, and application. Increased phenotyping capacity is necessary to usher in the next generation of improved crops for a burgeoning population.

## DATA AVAILABILITY

Source code for Prospector can be found on GitHub: https://github.com/PhenoApps/Prospector

Prospector can be found in the Google Play Store:

https://play.google.com/store/apps/details?id=org.phenoapps.prospector

## CONFLICT OF INTEREST

None declared.

## ACKNOWLEDGEMENTS

Many thanks to the IITA Cassava breeding team and the NextGen Cassava Project for testing Prospector in their breeding program.

## FUNDING

The development of Prospector was supported by the National Science Foundation under Grant No. (1543958). This app is made possible by the support of the American People provided to the Feed the Future Innovation Lab for Crop Improvement through the United States Agency for International Development (USAID). The contents are the sole responsibility of the authors and do not necessarily reflect the views of USAID or the United States Government.

Program activities are funded by the United States Agency for International Development (USAID) under Cooperative Agreement No. 7200AA-19LE-00005.

# FIGURES

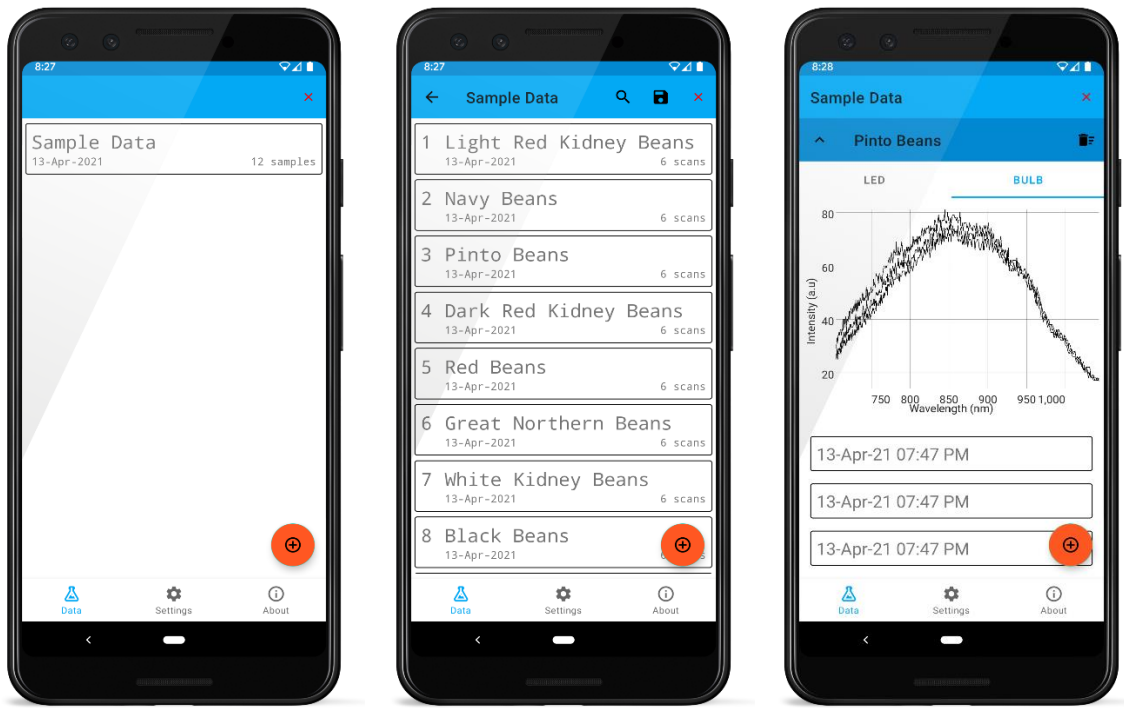

*Figure 1. Left: The list of current experiments stored in Prospector. Center: The list of samples within a selected experiment (e.g. Sample Data). Right: The list of scans within a selected sample (e.g. Pinto Beans).*

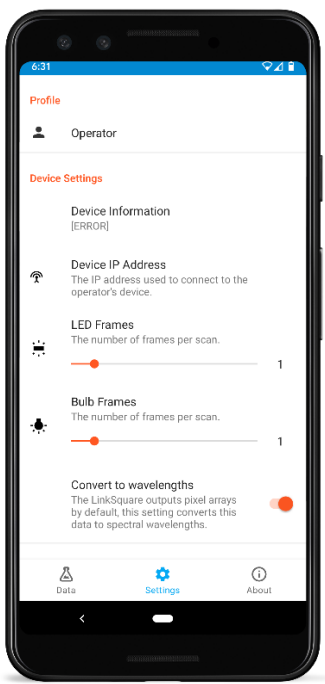

*Figure 2. Settings in Prospector.*